\documentclass[aps,pra,showpacs,twocolumn,superscriptaddress]{revtex4}%
\usepackage{bm}
\usepackage{graphics}
\usepackage{amsmath}
\usepackage{graphicx}
\usepackage{amsfonts}
\usepackage{amssymb}%
\begin{document}
\title{Gaussian two-mode attacks in one-way quantum cryptography}

\pacs{03.67.Dd, 03.65.-w, 42.50.-p, 89.70.Cf}
\author{Carlo Ottaviani}
\email{carlo.ottaviani@york.ac.uk}
\affiliation{Department of Computer Science \& York Center for Quantum Technologies,
University of York, York YO10 5GH, United Kingdom}
\author{Stefano Mancini}
\affiliation{School of Science and Technology, University of Camerino, Camerino (MC),
I-62032, Italy}
\affiliation{INFN Sezione di Perugia, I-61023, Perugia, Italy}
\author{Stefano Pirandola}
\affiliation{Department of Computer Science \& York Center for Quantum Technologies,
University of York, York YO10 5GH, United Kingdom}

\begin{abstract}
We investigate the asymptotic security of one-way continuous variable quantum
key distribution against Gaussian two-mode coherent attacks. The one-way
protocol is implemented by arranging the channel uses in two-mode blocks. By
applying symmetric random permutations over these blocks, the security
analysis is in fact reduced to study two-mode coherent attacks and, in
particular, Gaussian ones, due to the extremality of Gaussian states. We
explicitly show that the use of two-mode Gaussian correlations by an
eavesdropper leads to asymptotic secret key rates which are \textit{strictly}
larger than the rate obtained under standard single-mode Gaussian attacks.

\end{abstract}
\maketitle

\section{Introduction}

Quantum technologies are becoming reality, with huge efforts being devoted to
developing scalable quantum computers and robust quantum communications, e.g.,
for building a future quantum
Internet~\cite{Kimble2008,HybridINTERNET,telereview,Ulrikreview,KurizkiPNAS15}%
. In this global scenario, quantum key distribution
(QKD)~\cite{SCARANI,GISIN-RMP,diamanti2015} is certainly one of the most
advanced areas, with intense research activities directed towards practical
implementations. QKD represents a set of strategies that, integrating both
quantum and classical communication, allow two authorized remote users (Alice
and Bob) to generate a random sequence of bits; this is then used as an
encryption key in a one-time pad protocol~\cite{CLASSCRYPTO}, therefore
providing an unconditionally secure (information-theoretic~\cite{SHANNON49})
private communication between the remote users.

The effectiveness of QKD relies on the ground rule of encoding classical
information in non-orthogonal quantum states~\cite{BB84}, that are then
transmitted through a noisy quantum channel controlled by the eavesdropper
(Eve). This is also equivalent to sending the \textquotedblleft non-orthogonal
part\textquotedblright\ of discordant quantum states~\cite{DiscordQKD}. In
this way, Eve's attack is bounded by fundamental laws of quantum
physics~\cite{nocloning}: Any information gained by Eve creates loss and noise
on the quantum channel. Thanks to this trade-off, Alice and Bob can accurately
quantify the amount of classical error correction and privacy amplification
needed to reduce Eve's stolen information to a negligible
amount~\cite{GISIN-RMP}.

Since the first proposals to implement quantum information and computational
tasks, continuous variable (CV) systems have attracted increasing
attention~\cite{SAM-RMP,RMP}. The fact of using quantum systems with
continuous spectra (infinite-dimensional Hilbert spaces) has several
advantages with respect to the traditional approach based on discrete
variables (qubits). In particular, one can implement QKD at \textit{high
rates} by using highly-modulated coherent states and homodyne detections, not
only in one-way
schemes~\cite{grosshansNAT,WEEDNOSWITCH,jouguetNATPHOT,WEEDTH,Usenko10,thermal-PRA}%
, but also in two-way
protocols~\cite{Twoway,WEEDTH2,2way-ON,2WAY-COH,Two-wayGUO}
and\ CV\ strategies based on measurement-device independence
(MDI)~\cite{PIRS-RELAY,RELAY-SYMM,REACT,Joseph}

Ideal implementations of CV-QKD provide the highest key rates, not so far from
the ultimate repeaterless bound recently established in Ref.~\cite{stretching}%
. For a lossy channel with transmissivity $\tau$, the maximum rate achievable
by any QKD\ protocol (secret-key capacity) is equal to~\cite{stretching}
$K=-\log_{2}(1-\tau)$, with a fundamental rate-loss scaling of $\tau
/\ln2\simeq1.44\tau$ bits per channel use for long distances, i.e., at high
loss $\tau\simeq0$. The most practical one-way CV-QKD protocols, i.e., the
switching~\cite{grosshansNAT} and no-switching~\cite{WEEDNOSWITCH} protocols,
can potentially reach an asymptotic long-distance rate of $\tau/\ln4$ bits per
use, which is half the secret key capacity. Similar performance for CV-MDI-QKD
in the most asymmetric configuration~\cite{gae-SPIE}.

In this work we deepen the study of the secret key rates of the most known
one-way CV-QKD protocols~\cite{grosshansNAT,WEEDNOSWITCH}. In particular, we
explicitly study their security in the presence of Gaussian two-mode attacks,
representing the residual eavesdropping strategy after the de Finetti
symmetrization~\cite{renner-2007,renner-cirac} over two-mode blocks. Under
these attacks, we derive the analytical expressions of the asymptotic key
rates~\cite{devetak}. With these in hands, we show that eavesdropping
strategies based on correlated ancillas turn out to be \textit{strictly} less
effective than Gaussian attacks based on uncorrelated ancillas (single-mode
attacks). In other words, any two-mode Gaussian attack with strictly non-zero
correlations improves Alice and Bob's key rate.\begin{figure}[ptb]
\vspace{-0.0cm}
\par
\begin{center}
\includegraphics[width=0.35\textwidth]{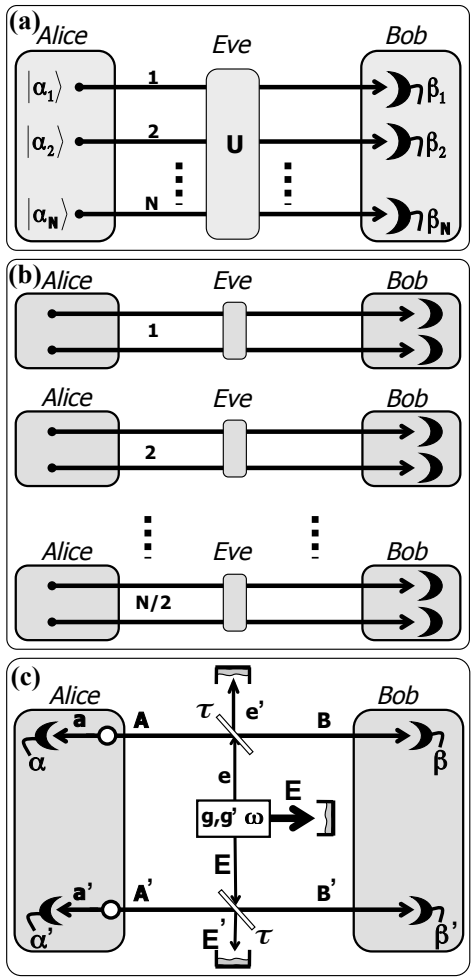}
\end{center}
\par
\vspace{-0.0cm}\caption[One-way general scheme]{Reduction to Gaussian two-mode
attacks. \textbf{(a)}~Alice Gaussianly modulates $N$ coherent states
$|\alpha_{k}\rangle$ in an independent and identical fashion. These are sent
through a quantum channel (Eve) and received by Bob, whose measurements
provide the classical variables $\beta_{k}$ for $k=1,...,N$. Eve's general
eavesdropping is based on a global unitary operation, $U$, applied to the $N$
instances of the one-way communication. \textbf{(b)}~After random
permutations, the coherence of the general attack is confined within each
two-mode block. \textbf{(c)}~Within an arbitrary block, we show a Gaussian
two-mode attack against the protocol (in EB representation). A realistic
Gaussian attack is simulated by two beam splitters, with transmissivity $\tau
$, mixing Alice's signals, $A$ and $A^{\prime}$, with Eve's ancillary modes,
$e$ and $E$, belonging to a larger set of modes $\{e,E,\mathbf{E}%
^{\prime\prime}\}$ in her hands. The reduced state of modes $e$ and $E$ is
Gaussian with thermal noise $\omega$ and correlation matrix $\mathbf{G}$ as in
Eq.~(\ref{VEVE}).}%
\label{scheme}%
\end{figure}

\section{Protocol and general considerations\label{section2}}

Let us consider the communication scheme of Fig.~\ref{scheme}(a). Alice sends
to Bob $N\gg1$ coherent states $|\alpha_{k}\rangle$. The amplitudes
$\alpha_{k}$, for $k=1,...,N$, are independently and identically modulated by
a bivariate zero mean Gaussian distribution of variance $\mu$. The
communication channel is under Eve's control, and the output detections
provide Bob with classical outcomes ${\beta}_{k}${.} After $N$ uses of the
channel, the parties share two correlated random sequences of symbols given by
the sets $\{\alpha_{k}\}$ and $\{\beta_{k}\}$.

For the sake of clarity, we consider reverse reconciliation (RR), so that the
key is obtained by Alice inferring Bob's variables. Now, when Bob\ applies
homodyne\emph{ }detections, randomly switching between measurements on
quadrature $\hat{q}_{k}$ and $\hat{p}_{k}$, we have the switching
protocol~\cite{grosshansNAT}. By contrast, when Bob measures both quadratures
(heterodyne detection), we have the no-switching protocol~\cite{WEEDNOSWITCH}.
Here we discuss the latter case, while we leave the analysis of the switching
protocol in Appendix \ref{APP1}.

In a general attack, Eve applies a global unitary operation $U$, which
coherently process her ancillary modes with all the $N$ signals exchanged by
the parties, with the ancillary outputs stored in a quantum memory. One has
that Bob-Eve joint system is described by a quantum state in the following
form%
\begin{equation}
\rho=U(\bigotimes_{k=1}^{N}|\alpha_{k}\rangle\langle\alpha_{k}|\otimes
|\Phi\rangle_{Eve}\langle\Phi|)U^{\dag}, \label{Qc}%
\end{equation}
where $|\Phi\rangle_{Eve}$ is Eve's total input state. The security analysis
considering this general scenario is not a practically solvable problem but,
in the limit of $N\rightarrow\infty$, it has been proved
\cite{renner-2007,renner-cirac} that one can get rid of the cross-correlations
between different uses of the channel. More specifically, with no loss of
generality, the security analysis can be simplified by applying symmetric
random permutations on the input ($\{\alpha_{k}\}$) and output ($\{\beta
_{k}\}$)\ classical data-sets.

Note that Alice and Bob may arrange the signals into two-mode blocks $c_{j}$,
with $j=1,...,N/2$. Then, they can apply random permutations over the blocks
$c_{j}$ rather than over the single uses of the channel. After such a
symmetrization, the quantum state given in Eq.~(\ref{Qc}) can be rewritten as
the following tensor product%
\begin{equation}
\rho\simeq\bigotimes_{j=1}^{M}\rho_{block},
\end{equation}
where $M=N/2$ is large. After this symmetrization, the initial global
coherence of quantum state of Eq.~(\ref{Qc}) is reduced to that one enclosed
within each two-mode state $\rho_{block}$, associated with the arbitrary block
$c_{j}$, as also depicted in Fig.~\ref{scheme} (b). Thus, the only effective
coherence to consider is two-mode and this scenario can be further simplified
using the extremality of Gaussian states~\cite{Wolf}.

In other words, the previous assumptions allow us to reduce the general
eavesdropping strategy to a Gaussian two-mode attack within each block. In
particular, we may consider the most realistic form of such an attack, where
Eve exploits two beam splitters to combine Alice's signals with correlated
ancillas prepared in an arbitrary Gaussian state. See Fig.~\ref{scheme}(c).
Note that this is a reduction which is often considered in practice.
The\ security analysis of one-way CV-QKD protocols under collective
(single-mode) Gaussian attacks~\cite{attacks} is typically restricted to the
most practical case of entangling-cloner attacks, resulting in thermal-loss
channels between Alice and Bob. The optimal key rate achievable over this
channel has been recently upper-bounded in Ref.~\cite{stretching} and
lower-bounded in Ref.~\cite{thermalBOUND}.

\section{Entanglement-based representation and Gaussian two-mode attacks}

The security analysis is performed in the entanglement based (EB)
representation~\cite{grosshansEB,RMP}, as also shown in Fig.~\ref{scheme}(c).
Alice owns a source of two-mode squeezed vacuum (TMSV) states. These are
zero-mean Gaussian states with covariance matrix (CM) of the form%
\begin{equation}
\mathbf{V}_{EPR}=\left(
\begin{array}
[c]{cc}%
\mu\mathbf{I} & \sqrt{\mu^{2}-1}\mathbf{Z}\\
\sqrt{\mu^{2}-1}\mathbf{Z} & \mu\mathbf{I}%
\end{array}
\right)  , \label{EPR}%
\end{equation}
where $\mu\geqslant1$, $\mathbf{I}=$\textrm{diag}$(1,1)$ and $\mathbf{Z}%
=$\textrm{diag}$(1,-1)$. In each block, Alice's input state is Gaussian of the
form $\rho_{aA}\otimes\rho_{a^{\prime}A^{\prime}}$ and CM $\mathbf{V}%
_{EPR}\oplus\mathbf{V}_{EPR}$. The signal coherent states $|\alpha\rangle$ and
$|\alpha^{\prime}\rangle$ are remotely projected on modes, $A$ and $A^{\prime
}$, by applying heterodyne detections on local modes $a$ and $a^{\prime}$. In
this way Alice modulates the amplitudes $\alpha$ and $\alpha^{\prime}$
according to a zero-mean Gaussian distribution with variance $\mu-1$ (which is
typically large).

As previously mentioned, we assume a realistic Gaussian two-mode attack where
Eve employs two identical beam-splitters with transmissivity $\tau$. These are
used to mix Alice's input modes, $A$ and $A^{\prime}$, with Eve's ancillary
modes, $e$ and $E$, respectively. The latter belong to a larger set of
ancillary states $\{e,E,\mathbf{E}^{\prime\prime}\}$ owned by the
eavesdropper. The reduced Gaussian state $\sigma_{eE}$ is completely
determined by the following CM~\cite{NJP2013}%
\begin{equation}
\mathbf{V}_{eE}=\left(
\begin{array}
[c]{cc}%
\omega\mathbf{I} & \mathbf{G}\\
\mathbf{G} & \omega\mathbf{I}%
\end{array}
\right)  ,\text{ for }\mathbf{G}:=\left(
\begin{array}
[c]{cc}%
g & 0\\
0 & g^{\prime}%
\end{array}
\right)  , \label{VEVE}%
\end{equation}
where $\omega=2\bar{n}+1$ quantifies Eve's thermal noise, with $\bar{n}$ mean
number of thermal photons. The correlations between modes $e$ and $E$ are
described by the parameters $g$ and $g^{\prime}$ in the matrix $\mathbf{G}$.
Their values are bounded\ by the constraints
\begin{equation}
|g|<\omega,|g^{\prime}|<\omega\text{, and }\omega\left\vert g+g^{\prime
}\right\vert \leq\omega^{2}+gg^{\prime}-1, \label{CONST}%
\end{equation}
which are imposed by the the uncertainty principle \cite{NJP2013,PIRS-RELAY}.
Note that from the CM of Eq.~(\ref{VEVE}), one can recover the standard
collective attack scenario (single mode attack) for $g=g^{\prime}=0$.

In the ideal case of perfect RR\ efficiency, the key-rate (bit per channel
use) is defined as%
\begin{equation}
R=\frac{I_{AB}-I_{E}}{2}, \label{KEYRATE1}%
\end{equation}
where $I_{AB}$ is the mutual information between variables $\{\alpha,\beta\}$
and $\{\alpha^{\prime},\beta^{\prime}\}$ and $I_{E}$ is Eve's accessible
information on Bob's variables (factor $2$ accounts for the double use of the
channel within each block). For many uses of the channel $N\gg1$, $I_{E}$ is
bounded by the Holevo information
\begin{equation}
\chi=S_{E}-S_{E|\beta\beta^{\prime}}=S_{AB}-S_{A|\beta\beta^{\prime}}.
\label{HOLEVO-DEF}%
\end{equation}
Here $S_{E}$ is the entropy of Eve's reduced state $\rho_{E}$ which is equal
to the entropy $S_{AB}$ of Alice and Bob's joint state $\rho_{AB}%
=\rho_{aa^{\prime}BB^{\prime}}$ (because the global state of Alice, Bob and
Eve is pure). Then, $S_{E|\beta\beta^{\prime}}$ is the entropy of Eve's state
$\rho_{E|\beta\beta^{\prime}}$\ conditioned on Bob variables $\beta$ and
$\beta^{\prime}$; Because these are the outcomes of a rank-1 measurement, we
have that Alice's conditional state $\rho_{A|\beta\beta^{\prime}}$ has entropy
$S_{A|\beta\beta^{\prime}}=S_{E|\beta\beta^{\prime}}$.\ 

Nore that, for Gaussian states, the von Neumann entropy $S(.)$ can be computed
via the formula%
\begin{equation}
S:=%
{\textstyle\sum\nolimits_{x}}
h(x), \label{S-GEN}%
\end{equation}
where $x$ are symplectic eigenvalues~\cite{RMP} and%
\begin{equation}
h(x):=\frac{x+1}{2}\log_{2}\frac{x+1}{2}-\frac{x-1}{2}\log_{2}\frac{x-1}{2},
\label{h-func}%
\end{equation}
By replacing $I_{E}$ in Eq.~(\ref{KEYRATE1}) with the Holevo function of
Eq.~(\ref{HOLEVO-DEF}), one obtains the following ideal key-rate (in RR)%
\begin{equation}
R:=\frac{I_{AB}-\chi}{2}. \label{KEY-RATE-teo}%
\end{equation}

\section{Security analysis}

\subsection{Mutual Information}

As a consequence of the two-mode reduction strategy, Alice and Bob's mutual
information is given by%
\begin{equation}
I_{AB}=I+I^{\prime}, \label{IAB1}%
\end{equation}
where $I:=I(\alpha,\beta)$ is the contribution from the first channel use, and
$I^{\prime}:=I(\alpha^{\prime},\beta^{\prime})$ from the second use. Each
contribution is given by the following expression%
\begin{equation}
I^{(\prime)}=\log_{2}\frac{V_{B}+1}{V_{B|\alpha(\alpha^{\prime})}+1},
\label{IAB2}%
\end{equation}
where $V_{B}=\tau\mu+(1-\tau)\omega$ describes the quadrature variance of the
average thermal state arriving at Bob's side, while $V_{B|\alpha}%
=V_{B|\alpha^{\prime}}=\tau+(1-\tau)\omega$ is the quadrature variance of
Bob's state after Alice's heterodyne detection. Using these relations in
Eqs.~(\ref{IAB1}) and~(\ref{IAB2}), and working in the limit of $\mu\gg1$, one
easily obtains%
\begin{equation}
I_{AB}=2\log_{2}\frac{\tau\mu}{1+\tau+(1-\tau)\omega}. \label{IAB3}%
\end{equation}
We note that, as one would expect, this expression does not depend on the
correlation parameters $g$ and $g^{\prime}$.

\subsection{Holevo Bound}

We now describe the general steps to obtain the Holevo bound $\chi$ (more
details are in Appendix~\ref{APP-crypto}).\ Working in the EB representation,
Alice and Bob's joint state $\rho_{aa^{\prime}BB^{\prime}}$ is described by
the following CM%
\begin{equation}
\mathbf{V}=\left(
\begin{array}
[c]{cccc}%
(\mu+1)\mathbf{I} &  & \Phi\mathbf{Z} & \\
& (\mu+1)\mathbf{I} &  & \Phi\mathbf{Z}\\
\Phi\mathbf{Z} &  & \Lambda\mathbf{I} & (1-\tau)\mathbf{G}\\
& \Phi\mathbf{Z} & (1-\tau)\mathbf{G} & \Lambda\mathbf{I}%
\end{array}
\right)  , \label{VTOT-text}%
\end{equation}
where we have set
\begin{align}
\Lambda &  :=\tau(\mu+1)+(1-\tau)\omega,\label{LAMBDA}\\
\Phi &  :=\sqrt{\tau\lbrack(\mu+1)^{2}-1]}, \label{la2}%
\end{align}
The symplectic spectrum is obtained from the ordinary eigenvalues of matrix
$|i\Omega\mathbf{V}_{tot}|$~\cite{RMP} with
\begin{equation}
\Omega=\mathbf{\omega\oplus\omega},~~\mathbf{\omega=}\left(
\begin{array}
[c]{cc}%
0 & 1\\
-1 & 0
\end{array}
\right)  ~.
\end{equation}

In the limit of large $\mu$, and after some simple algebra, we find the
following symplectic eigenvalues%
\begin{align}
\nu_{+}  &  =\sqrt{(\omega+g)(\omega+g^{\prime})},\\
\nu_{-}  &  =\sqrt{(\omega-g)(\omega-g^{\prime})},\\
\nu_{1}  &  =\nu_{2}=(1-\tau)\mu.
\end{align}
Using these eigenvalues and the expansion%
\begin{equation}
h(x)\simeq\log_{2}\frac{e}{2}x+O\left(  x^{-1}\right)  , \label{limH}%
\end{equation}
we find the following expression for Alice and Bob's von Neumann entropy%
\begin{equation}
S_{AB}=h(\nu_{+})+h(\nu_{-})+2\log_{2}\frac{e}{2}(1-\tau)\mu. \label{vonQQQ}%
\end{equation}

The next step is to apply two sequential heterodyne detections on modes $B$
and $B^{\prime}$, to obtain the conditional CM $\mathbf{V}_{C}$ describing the
conditional quantum state $\rho_{aa^{\prime}|\beta\beta^{\prime}}$. The
corresponding CM has a complicated expression that can be found in
Eq.~(\ref{VC-noswitch}) of Appendix~\ref{APP-crypto}. Computing its symplectic
eigenvalues in the limit of $\mu\gg1$, we find the following conditional
spectrum%
\begin{equation}
\{\bar{\nu}_{+},\bar{\nu}_{-}\}=\left\{  \frac{\sqrt{\lambda_{+}\lambda
_{+}^{\prime}}}{\tau},\frac{\sqrt{\lambda_{-}\lambda_{-}^{\prime}}}{\tau
}\right\}  , \label{SPECTRUM-COND-HET2}%
\end{equation}
where we have defined
\begin{align}
\lambda_{\pm}  &  :=1+(1-\tau)(\omega\pm g),\\
\lambda_{\pm}^{\prime}  &  :=1+(1-\tau)(\omega\pm g^{\prime}).
\end{align}
The conditional entropy just reads%
\begin{equation}
S_{A|\beta\beta^{\prime}}=h(\bar{\nu}_{+})+h(\bar{\nu}_{-}).
\label{TOTAL-vonNEUMANNHETHET}%
\end{equation}
Finally, using Eqs.~(\ref{vonQQQ}) and (\ref{TOTAL-vonNEUMANNHETHET}) in
Eq.~(\ref{HOLEVO-DEF}), we can write Eve's Holevo bound as%
\begin{equation}
\chi=2\log_{2}\frac{e}{2}(1-\tau)\mu+\sum_{i=\pm}\left[  h(\nu_{i})-h(\bar
{\nu}_{i})\right]  . \label{Holevo}%
\end{equation}
It is easy to check that Eq.~(\ref{Holevo}) recovers the expression of the
Holevo bound of standard collective (single-mode) Gaussian attacks for
$g=g^{\prime}=0$.

\subsection{Secret key rate and its analysis}

It is easy to compute the secret-key rate using Eq.~(\ref{IAB3}) and
(\ref{Holevo}) in Eq.~(\ref{KEY-RATE-teo}). After some algebra, we obtain the
following expression for the rate of the no-switching protocol under realistic
Gaussian two-mode attacks%
\begin{align}
R  &  =\log_{2}\frac{2}{e}\frac{\tau}{(1-\tau)[1+\tau+(1-\tau)\omega
]}\nonumber\\
&  +\frac{1}{2}\sum_{i=\pm}\left[  h(\bar{\nu}_{i})-h(\nu_{i})\right]  .
\label{RateHET2}%
\end{align}
In order to prove that Gaussian two-mode attacks with non-zero correlations
are strictly less effective than single-mode attacks, we study the derivatives
of this rate. We find the following strict inequality%
\begin{equation}
R(\tau,\omega,g,g^{\prime})>R(\tau,\omega,0,0),\text{ \ \ }\forall
g,g^{\prime}\neq0. \label{R-Rcoll}%
\end{equation}

The details of the proof are in Appendix \ref{APP-crypto}, while here we limit
the discussion to the general ideas. To show Eq.~(\ref{R-Rcoll}), we first
seek for critical points of the function $R(\tau,\omega,g,g^{\prime})$.
Solving the equation $\nabla R=0$ on the $(g,g^{\prime})$-plane, one finds
that only the origin $P_{0}:=(0,0)$ is critical. To determine the nature of
$P_{0}$, we then compute the second-order derivatives with respect the
correlation parameters $g$ and $g^{\prime}$. This allows us to compute the
expression of the Hessian matrix $H$ and study its positive definiteness. We
therefore find that $P_{0}$ corresponds to the absolute minimum of the rate in
Eq.~(\ref{RateHET2}) within the domain defined by Eq.~(\ref{CONST}).

\begin{figure}[ptb]
\vspace{-0.0cm}
\par
\begin{center}
\includegraphics[width=0.52\textwidth]{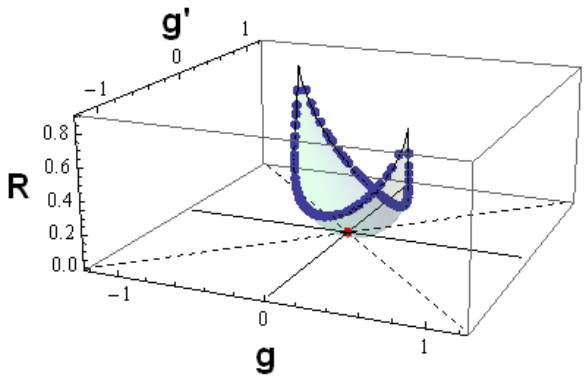}
\end{center}
\par
\vspace{-0.0cm}\caption[One-way general scheme]{We analyze the key-rate of Eq.
(\ref{RateHET2}) over the plane of the correlation parameters, $g$ and
$g^{\prime}$. Any two-mode attack corresponds to a point in the colored
surface. Boundary attacks, verifying the condition $\omega|g+g^{\prime
}|=\omega^{2}+gg^{\prime}-1$, are represented by the blue points. The rate of
the single-mode attack $g=g^{\prime}=0$ is the red spot. Here we fix
$\tau\simeq0.44$ and $\omega=1.2$. For these values, the single-mode attack
provides zero key-rate. On the other hand, we see that the key rate is
positive for any two-mode attack with non-zero correlations.}%
\label{RATE-boundaries}%
\end{figure}

Finally we check that the attacks over the boundary, given by the condition
$\omega\left\vert g+g^{\prime}\right\vert =\omega^{2}+gg^{\prime}-1$, also
provide key rates which are strictly larger than that under the single-mode
attack. In Fig.~\ref{RATE-boundaries} we show a numerical example, which is
obtained by fixing the transmissivity $\tau\simeq0.44$, the thermal noise
$\omega=1.2$, and plotting the rate as a function of $g$ and $g^{\prime}$. We
see that the secret-key rate under single-mode attack (red dot) is always
strictly less than that the rate which is obtained by any physically-permitted
two-mode attack (which is a point in the colored surface). The key rates for
the attacks on the boundary of this region are the blue dots.
The origin $P_{0}$ is therefore always an absolute minimum for
$R$. As a consequence, any correlation injected into the channels by the
eavesdropper to implement the coherent attack automatically increases the key rate.

\section{Conclusion}

In this work we have explicitly studied the security of one-way CV-QKD
protocols against Gaussian two-mode attacks. The approach is based on an
attack-reduction strategy where the parties pack the uses of the quantum
channel in two-mode blocks. Then, they apply random permutations over these
blocks. This allows them to get rid of any cross correlation engineered by the
eavesdropper between different blocks. We solved this problem analytically,
and we obtained the secret-key rates under Gaussian two-mode attacks, in
particular, those more realistic and based on a suitable combination of
entangling cloners.

We have then showed that any non-zero correlation used by the eavesdropper
leads to a strictly higher key-rate than the rate obtained under Gaussian
single-mode attacks. This is achieved under the condition that infinite
signals are exchanged (asymptotic rate), therefore not considering composable
or finite-size analyses~\cite{leverrier2015}. We conjecture that the use of
correlations is not effective even when the size of the blocks is greater than
two modes. It would be interesting to check if this is still true if Alice
adopted correlated encodings between different uses of the
channel~\cite{ruggeri-mancini}.

\section{Acknowledgements}

This work has been supported by the EPSRC via the `UK Quantum Communications
HUB' (Grant no. EP/M013472/1).


\appendix

\section{Computations for the no-switching protocol\label{APP-crypto}}

Here we provide the calculations to prove Eq.~(\ref{R-Rcoll}) for the
no-switching protocol.

\subsection{Total covariance matrix}

Let $\mathbf{X=(}\hat{q}_{X},\hat{p}_{X}\mathbf{)}$ be the vectorial
quadrature operator describing a general mode $X$. The impact of the
attenuation and noise on the Alice's modes, $A$ and $A^{\prime}$, through two
identical beam splitters of transmissivity $\tau$ are given by the following
expressions%
\begin{align}
\mathbf{B}  &  =\sqrt{\tau}\mathbf{A}+\sqrt{1-\tau}\mathbf{e,}\label{B}\\
\mathbf{B}^{\prime}  &  =\sqrt{\tau}\mathbf{A}^{\prime}+\sqrt{1-\tau
}\mathbf{E,} \label{Bprime}%
\end{align}
where $\mathbf{e}$\textbf{ }and $\mathbf{E}$ are the vectorial quadrature
operators describing Eve's ancillary modes, $e$ and $E$, mixed at the beam
splitters with modes $A$ and $A^{\prime}$, respectively. Eve's reduced state
$\sigma_{eE}$\ is zero-mean Gaussian with CM\ as in Eq.~(\ref{VEVE}), with
local thermal noise $\omega$\ and correlation parameters $\mathbf{G}$
$:=\mathrm{diag}(g,g^{\prime})$ fulfilling the constraints of Eq.~(\ref{CONST}%
). We order Alice and Bob's output modes as follows $a,a^{\prime},B,B^{\prime
}$; then, we use Eqs.~(\ref{B}) and~(\ref{Bprime}) to compute the CM
describing Alice and Bob's total state $\rho_{aa^{\prime}BB^{\prime}}$. It is
simple to derive the following expression%
\begin{equation}
\mathbf{V}=\left(
\begin{array}
[c]{cccc}%
(\mu+1)\mathbf{I} &  & \Phi\mathbf{Z} & \\
& (\mu+1)\mathbf{I} &  & \Phi\mathbf{Z}\\
\Phi\mathbf{Z} &  & \Lambda\mathbf{I} & (1-\tau)\mathbf{G}\\
& \Phi\mathbf{Z} & (1-\tau)\mathbf{G} & \Lambda\mathbf{I}%
\end{array}
\right)  , \label{VTOT}%
\end{equation}
where $\mu-1$ is the classical Gaussian modulation, while $\Lambda$ and $\Phi$
are defined in Eqs.~(\ref{LAMBDA}) and~(\ref{la2}).

\subsection{Alice and Bob's mutual information}

In the no-switching protocol, Bob performs heterodyne detections measuring
both quadratures $\hat{q}$ and $\hat{p}$. From the form of the attack, we have
that the\ variances in $\hat{q}$ and $\hat{p}$, relative to both Bob's modes
$B$ and $B^{\prime}$, are identical and given by $V_{B}=\Lambda$, with
$\Lambda$ specified in Eq.~(\ref{LAMBDA}). The conditional variances, after
Alice's heterodyne detections, are given by%
\begin{equation}
V_{B|\alpha,\alpha^{\prime}}=\tau+(1-\tau)\omega.
\end{equation}
Accounting for the double use of the channel within the block, we derive the
mutual information%
\begin{equation}
I_{AB}=2\log_{2}\frac{V_{B}+1}{V_{B|\alpha,\alpha^{\prime}}+1}.
\end{equation}
Taking the limit of large modulation ($\mu\gg1$), one gets the asymptotic
expression of the mutual information, given in Eq.~(\ref{IAB3}) of the main
text, i.e.,%
\begin{align}
I_{AB}  &  =2\log_{2}\frac{\tau(\mu+1)+(1-\tau)\omega+1}{1+\tau+(1-\tau
)\omega}\nonumber\\
&  \overset{\mu\rightarrow\infty}{\rightarrow}2\log_{2}\frac{\tau\mu}%
{1+\tau+(1-\tau)\omega}. \label{IABAPP}%
\end{align}

\subsection{Computation of the Holevo bound}

The EB representation and dilation of the two-mode channel allows us to
describe the joint Alice-Bob-Eve\ output state as pure. Noting that this
quantum state is always processed by rank-$1$ measurements, one has that the
purity is also preserved on the conditional state after detection. The
eavesdropper is assumed to control the quantum memory storing her ancillary
modes, she is computationally unbounded, but the parties exchange an infinite
number of signals, $N\gg1$. In this regime Eve's accessible information
$I_{E}$ on Bob's variables is bounded by the Holevo quantity $\chi$. It can be
obtained from the von Neumann entropy of Alice-Bob total state $S(\rho
_{aa^{\prime}BB^{\prime}})$, and the conditional von Neumann entropy
$S(\rho_{aa^{\prime}|\beta\beta^{\prime}})$. The Holevo bound is then given
by
\begin{equation}
\chi=S(\rho_{aa^{\prime}BB^{\prime}})-S(\rho_{aa^{\prime}|\beta\beta^{\prime}%
}).
\end{equation}

We need to derive the function $\chi$ in terms of the relevant parameters of
the protocol $\tau$, $\omega$, $g$, and $g^{\prime}$. We then compute the
symplectic spectrum of the total CM given by Eq.~(\ref{VTOT}), from the
absolute value of the eigenvalues of the matrix $\mathbf{M}%
=i\bm{\Omega}\mathbf{V}_{tot}$, where $\mathbf{\Omega}=\oplus_{k=1}%
^{4}\mathbf{\omega}$ is the $8\times8$ (four modes) symplectic form
\cite{RMP}. For large $\mu$, one obtains the following expressions%
\begin{align}
\nu_{+}  &  =\sqrt{(\omega+g)(\omega+g^{\prime})},\label{SPECTRUMTOT1}\\
\nu_{-}  &  =\sqrt{(\omega-g)(\omega-g^{\prime})},\label{SPECTRUMTOT2}\\
\nu_{1}  &  =\nu_{2}=(1-\tau)\mu, \label{SPECTRUMTOT3}%
\end{align}
which, together with Eq.~(\ref{S-GEN}) and Eq.~(\ref{limH}), are used to
calculate the total von Neumann entropy $S(\rho_{aa^{\prime}BB^{\prime}%
})=S_{AB}$ given in Eq.~(\ref{vonQQQ}).

Now, the conditional CM $\mathbf{V}_{C}$, providing the conditional von
Neumann entropy, is obtained via heterodyning Bob's modes $B$ and $B^{\prime}%
$. We apply the formula for heterodyne detection~\cite{OPENsys} to the total
CM\ $\mathbf{V}$. After some algebra, $\mathbf{V}_{C}$ can be written in the
following form%
\begin{equation}
\mathbf{V}_{C}=\frac{1}{(\Lambda+1)^{2}-g^{2}(1-\tau)^{2}}\left(
\begin{array}
[c]{cccc}%
k &  & \tilde{k} & \\
& k^{\prime} &  & \tilde{k}^{\prime}\\
\tilde{k} &  & k & \\
& \tilde{k}^{\prime} &  & k^{\prime}%
\end{array}
\right)  , \label{VC-noswitch}%
\end{equation}
with the matrix entries defined as%
\begin{align}
k  &  :=(\mu+1)[g^{2}(1-\tau)^{2}+(\Lambda+1)\tilde{\Lambda}]+(\Lambda
+1)\tau,\\
\tilde{k}  &  :=-g(1-\tau)\tau\mu(\mu+2),\\
\tilde{\Lambda}  &  :=\Lambda-\tau,\\
k^{\prime}  &  :=k(g\rightarrow g^{\prime}),\\
\tilde{k}^{\prime}  &  :=\tilde{k}(g\rightarrow g^{\prime}).
\end{align}
For large $\mu$, the symplectic spectrum of the conditional CM $\mathbf{V}%
_{C}$ is given by Eq.~(\ref{SPECTRUM-COND-HET2}). Note that this spectrum does
not depend on the modulation $\mu$, and for $g=g^{\prime}=0$ we recover the
conditional eigenvalues of Ref.~\cite{WEEDNOSWITCH}.

Now, from Eq.~(\ref{SPECTRUM-COND-HET2}), we derive the conditional von
Neumann entropy $S(\rho_{aa^{\prime}|\beta\beta^{\prime}})=S_{A|\beta
\beta^{\prime}}$ given in Eq.~(\ref{TOTAL-vonNEUMANNHETHET}). Combining the
computed entropies, we obtain the Holevo bound in Eq.~(\ref{Holevo}). Finally,
including the mutual information of Eq.~(\ref{IABAPP}), we derive the
asymptotic key rate%
\begin{align}
R_{\mathrm{Block}}  &  =\log_{2}\frac{4}{e^{2}}\frac{\tau^{2}}{(1-\tau
)^{2}[1+\tau+(1-\tau)\omega]^{2}}\label{RATE-FORMULA-HET2}\\
&  +\sum_{k=\pm}\left[  h(\bar{\nu}_{k})-h(\nu_{k})\right]  .\nonumber
\end{align}
More precisely, for channel use, we find%
\begin{equation}
R=\frac{R_{\mathrm{Block}}}{2},
\end{equation}
as given in Eq.~(\ref{RateHET2}).

\subsection{Study of the critical point}

From the first-order derivatives $\partial_{g}R$ and $\partial_{g^{\prime}}R$,
and solving the equation $\nabla R=0$, one finds a single critical point
$P_{0}$ for any $\tau$ and $\omega$; this is given by the origin
$(g=g^{\prime}=0)$ of the correlation plane ($g$, $g^{\prime}$), bounded by
the constraints given by Eq.~(\ref{CONST}). We then take the second-order
derivative $\partial^{2}R$, with respect to $g$ and $g^{\prime}$, and build
the (symmetric) Hessian matrix%
\begin{equation}
H=\left(
\begin{array}
[c]{cc}%
\partial_{g}^{2}R & \partial_{gg^{\prime}}^{2}R\\
\partial_{g^{\prime}g}^{2}R & \partial_{g^{\prime}}^{2}R
\end{array}
\right)  . \label{HESSIAN-MATRIX}%
\end{equation}
From the positive definiteness of this matrix, evaluated in the critical point
$P_{0}$, one has that $P_{0}$ is an absolute minimum. We then study the sign,
in $P_{0}$, of the determinant of the Hessian matrix~(\ref{HESSIAN-MATRIX}).

After some algebra one can write it in the simplified form%
\begin{equation}
\det H=\frac{D_{1}-D_{2}}{\tau\left[  \bar{\lambda}+\tau\right]  \bar{\lambda
}\omega(\omega^{2}-1)} \label{Hhethet}%
\end{equation}
where we have defined%
\begin{align}
f(x)  &  :=\frac{1}{\log_{2}e}\log_{2}\frac{1+x}{1-x}~~(>0\text{ for
}0<x<1)\label{func-f}\\
D_{1}  &  :=\tau\left[  f\left(  \omega^{-1}\right)  +2\log_{2}\frac
{\bar{\lambda}+\tau}{(1-\tau)\sqrt{\omega^{2}-1}}\right]  ,\\
D_{2}  &  :=\omega\left[  f\left(  \tau\bar{\lambda}^{-1}\right)  +\tau
^{2}\log_{2}\frac{\bar{\lambda}+\tau}{\bar{\lambda}+\tau-2}\right]  ,\\
\bar{\lambda}  &  :=1+\omega(1-\tau).
\end{align}
One can check that $f\left(  \tau\bar{\lambda}^{-1}\right)  \geq0$, and
$D_{1}>D_{2}$ for any $0\leq\tau\leq1$ and $\omega\geq1$. Indeed, being both
attenuation $\tau$ and noise $\omega$ positive quantities, as well as
$\bar{\lambda}$, we have%
\begin{equation}
\det H>0\text{~~for any}~\tau\text{ and}~\omega\text{.}%
\end{equation}

We then proceed with the study of the second-order derivative $\partial
_{g}^{2}R$ at the critical point $P_{0}$. This is the first principal minor of
the Hessian matrix of Eq.~(\ref{HESSIAN-MATRIX}). It is easy to check the
following chain of inequalities%
\begin{align}
\partial_{g}^{2}R  &  =\frac{1}{\left(  \tau+\bar{\lambda}\right)  (\omega
^{2}-1)}+\frac{f(\omega^{-1})}{4\omega}+\frac{(1-\tau)^{2}}{4\tau\bar{\lambda
}}f(\tau\bar{\lambda}^{-1})\nonumber\\
&  >\frac{1}{\left(  \tau+\bar{\lambda}\right)  (\omega^{2}-1)}+\frac
{f(\omega^{-1})}{4\omega}\nonumber\\
&  >\frac{1}{(\tau+\bar{\lambda})(\omega^{2}-1)}>0,~~\forall\omega>1\text{ and
}0\leq\tau\leq1 \label{D2RHET-g}%
\end{align}
Therefore, the extremal point $P_{0}$ is an absolute minimum for the key rate
of the no-switching protocol.

By contrast, we notice that the study described above is only valid for the
pairs ($g,$ $g^{\prime}$) for which it is possible to define the derivatives,
i.e., those lying within the domain bounded by the constraints of
Eq.~(\ref{CONST}). In order to complete our analysis we check that also the
points at the boundary of the domain, described by Eq.~(\ref{CONST}), give a
key rate which is larger than that one obtained for $g=g^{\prime}=0$. We have
studied numerically these cases, computing the rate for the pairs
$(g,g^{\prime})$ fulfilling the condition $\omega\left\vert g+g^{\prime
}\right\vert =\omega^{2}+gg^{\prime}-1$. In Fig.~\ref{RATE-boundaries} we show
an example of this computation, corresponding to the case of a transmissivity
$\tau\simeq0.44$ and thermal noise $\omega=1.3$, in shot-noise unit (SNU). We
see that the rate for single-mode collective attack (red spot) lies well below
the blue points, which describe the key rate for the boundary two-mode
attackes. The colored region gives the values of the\ key rate for any
non-zero correlations $g,g^{\prime}$.

Clearly, similar results are obtained for any other value of $0\leq\tau\leq1$
and $\omega\geq1$, with the area describing two-mode attacks vanishing into a
point as $\omega\rightarrow1$. In that case, the only possible attack is
single-mode and, according to Eq.~(\ref{CONST}), we have $g,g^{\prime
}\rightarrow0$.

\section{Switching\textit{ }protocol\label{APP1}}

In this section, we analyze the key rate and its critical point for the
switching protocol. We arrive at the same conclusion obtained for the
no-switching protocol.
In this case Bob performs homodyne detections on the received signals modes,
by randomly switching the quadratures measured. Within each block $c_{j}$, Bob
can decide to apply the same homodyne detection on both modes $B,B^{\prime}$,
or measure on two distinct bases ($\hat{q}$ and $\hat{p}$). Here we assume the
former case. When Bob detects both his modes in quadrature $\hat{q}$, we have%
\begin{align}
\mathbf{V}_{C}^{q} &  =\mu\mathbf{I-}\frac{\tau(\mu^{2}-1)}{\tilde{\Lambda
}[g^{2}(1-\tau)^{2}-\tilde{\Lambda}^{2}]}\nonumber\\
&  \times\left(
\begin{array}
[c]{cccc}%
2g^{2}(1-\tau)^{2}-\tilde{\Lambda}^{2} &  & g(1-\tau)\tilde{\Lambda} & \\
& 1 &  & \\
g(1-\tau)\tilde{\Lambda} &  & \tilde{\Lambda}^{2} & \\
&  &  & 1
\end{array}
\right)  ,
\end{align}
where $\tilde{\Lambda}=\tau\mu+(1-\tau)\omega=\Lambda-\tau$. When Bob detects
both his modes in quadrature $\hat{p}$, we obtain%
\begin{align}
\mathbf{V}_{C}^{p} &  =\mu\mathbf{I-}\frac{\tau(\mu^{2}-1)}{\tilde{\Lambda
}[g^{\prime2}(1-\tau)^{2}-\tilde{\Lambda}^{2}]}\nonumber\\
&  \times\left(
\begin{array}
[c]{cccc}%
1 &  &  & \\
& 2g^{\prime2}(1-\tau)^{2}-\tilde{\Lambda}^{2} &  & g^{\prime}(1-\tau
)\tilde{\Lambda}\\
&  & 1 & \\
& g^{\prime}(1-\tau)\tilde{\Lambda} &  & \tilde{\Lambda}^{2}%
\end{array}
\right)  .
\end{align}

In the first case ($\hat{q}$-detection), for large $\mu$, we obtain the
following symplectic spectrum%
\begin{equation}
\tilde{\nu}_{\pm}=\sqrt{\frac{(1-\tau)(\omega\pm g)\mu}{\tau}},
\label{SPECTR-COND1}%
\end{equation}
which depends on the correlation parameter $g$. In the second case ($\hat{p}%
$-detection), we have the following symplectic eigenvalues%
\begin{equation}
\tilde{\nu}_{\pm}^{\prime}=\sqrt{\frac{(1-\tau)(\omega\pm g^{\prime})\mu}%
{\tau}}, \label{SPECTR-COND2}%
\end{equation}
depending on correlation parameter $g^{\prime}$. From Eqs.~(\ref{SPECTR-COND1}%
) and~(\ref{SPECTR-COND2}), we compute two distinct conditional von Neumann
entropies,%
\begin{align}
S_{E|\beta_{q}\beta_{q}^{\prime}}  &  =h\left(  \tilde{\nu}_{+}\right)
+h\left(  \tilde{\nu}_{-}\right) \nonumber\\
&  \overset{\mu\rightarrow\infty}{=}\log_{2}\frac{e^{2}}{4}\frac{1-\tau}{\tau
}\sqrt{(\omega+g)(\omega-g)}\mu,
\end{align}
and%
\begin{align}
S_{E|\beta_{p}\beta_{p}^{\prime}}  &  =h\left(  \tilde{\nu}_{+}^{\prime
}\right)  +h\left(  \tilde{\nu}_{-}^{\prime}\right) \nonumber\\
&  \overset{\mu\rightarrow\infty}{=}\log_{2}\frac{e^{2}}{4}\frac{1-\tau}{\tau
}\sqrt{(\omega+g^{\prime})(\omega-g^{\prime})}\mu.
\end{align}
To the conditional von Neumann entropy, we average over these two cases,
getting the expression%
\begin{align}
S_{E|\beta\beta^{\prime}}  &  =\frac{S_{E|\beta_{q}\beta_{q}^{\prime}%
}+S_{E|\beta_{p}\beta_{p}^{\prime}}}{2}\nonumber\\
&  =\log_{2}\frac{e^{2}}{4}\frac{1-\tau}{\tau}\sqrt{\nu_{-}\nu_{+}}\mu.
\label{COND-vonNEUMANNHETHOM}%
\end{align}

\subsection{Key rate for the switching protocol}

Using the total von Neumann entropy of Eq.~(\ref{vonQQQ}), the conditional
entropy of Eq.~(\ref{COND-vonNEUMANNHETHOM}), and the asymptotic expression of
the mutual information for the switching protocol%
\begin{equation}
I_{AB}\rightarrow2\log_{2}\frac{\tau\mu}{\tau+(1-\tau)\omega},
\end{equation}
we compute the following expression of the key-rate against Gaussian two-mode
coherent attacks%
\begin{equation}
\tilde{R}=\frac{1}{2}\log_{2}\frac{\sqrt{\nu_{-}\nu_{+}}}{(1-\tau
)[\tau+(1-\tau)\omega]}-\frac{h\left(  \nu_{+}\right)  +h\left(  \nu
_{-}\right)  }{2}, \label{FORMULA-RATE-HET-HOM}%
\end{equation}
from which we can recover the standard case of single-mode collective attack
setting $g=g^{\prime}=0$.

For the sake of completeness, here we also discuss the case where Bob applies
different homodyne detections (one in $\hat{q}$, the other in $\hat{p}$),
within each two-mode block. In this case one finds a lower key rate because
measurements have the effect of de-correlating modes $B$ and $B^{\prime}$. As
a result, any dependency on $g,$ $g^{\prime}$ is cancelled from the
conditional CM, and for $\mu\gg1$ one finds the following doubly degenerate
eigenvalues%
\begin{equation}
\tilde{\nu}_{1,2}=\sqrt{\frac{(1-\tau)\omega\mu}{\tau}}. \label{MIX}%
\end{equation}
After some algebra we obtain the following non-optimal key rate%
\begin{equation}
\bar{R}=\frac{1}{2}\log_{2}\frac{\omega}{(1-\tau)[\tau+(1-\tau)\omega]}%
-\frac{h\left(  \nu_{+}\right)  +h\left(  \nu_{-}\right)  }{2},
\end{equation}
which is not interesting from a practical point of view, because the parties
can always choose to group instances of the protocol with the same quadrature homodyned.

\subsection{Study of the critical point for the switching protocol}

We then compute the first derivatives of the rate in
Eq.~(\ref{FORMULA-RATE-HET-HOM}), with respect to the correlations parameters
$g$ and $g^{\prime}$, obtaining the following%
\begin{align}
\partial_{g}\tilde{R} &  =\frac{\zeta}{4}\left[  f(\nu_{-}^{-1})+\frac
{g}{(\omega+g)\nu_{-}}-\frac{\nu_{+}\nu_{-}f(\nu_{+}^{-1})}{(\omega
+g)(\omega-g^{\prime})}\right]  \label{DRateDg}\\
\partial_{g^{\prime}}\tilde{R} &  =\frac{\zeta^{\prime}}{4}\left[  f(\nu
_{-}^{-1})+\frac{g^{\prime}}{(\omega+g^{\prime})\nu_{-}}-\frac{\nu_{+}\nu
_{-}f(\nu_{+}^{-1})}{(\omega+g^{\prime})(\omega-g)}\right]  ,\label{DRateDgp}%
\end{align}
where the function $f(.)$ has been defined in Eq.~(\ref{func-f}), and the
symplectic eigenvalues $\nu_{\pm}$ are given in Eqs. (\ref{SPECTRUMTOT1}) and
(\ref{SPECTRUMTOT2}), while we defined $\zeta$ and $\zeta^{\prime}$ as follows%
\begin{equation}
\zeta:=\frac{\nu_{-}}{2(\omega-g^{\prime})},\text{ }\zeta^{\prime}:=\frac
{\nu_{-}}{2(\omega-g)}.
\end{equation}
Note that these derivatives are properly defined within the constraints of
Eq.~(\ref{CONST}), that identify a sector of ($g,g^{\prime}$)- plane for which
the conditions $\nu_{-}>1$ and $\nu_{+}>1$ must hold. In fact, the situation
for which one has $\nu_{\pm}=1$ can only be obtained in $P_{0}$, i.e., if the
attack is collective. Solving the system of equations $\nabla R=0$ one finds
that $P_{0}$ is a critical point, and that it is also unique for any
$\omega\geq1$ and $g$ and $g^{\prime}$ fulfilling Eq.~(\ref{CONST}).

\subsection{Positive definiteness of the Hessian matrix}

The second-order derivatives with respect $g$, evaluated in $P_{0}$, is given
by%
\begin{align}
\partial_{g}^{2}\tilde{R}  &  =-\frac{\omega^{2}+g^{2}}{4\left(  \omega
^{2}-g^{2}\right)  ^{2}}+\frac{1}{8}\left[  \frac{\kappa_{+}}{\nu_{+}^{2}%
-1}-\frac{\kappa_{-}}{\nu_{-}^{2}-1}\right] \nonumber\\
&  +\frac{1}{8}\left[  \frac{\sqrt{\kappa_{+}}f(\nu_{+}^{-1})}{\omega+g}%
-\frac{\sqrt{\kappa_{-}}f(\nu_{-}^{-1})}{\omega-g}\right]
\end{align}
with the coefficients $\kappa_{\pm}$ defined as follows
\begin{equation}
\kappa_{+}:=\frac{\omega+g^{\prime}}{\omega+g},\text{\ \ }\kappa_{-}%
:=\frac{\omega-g^{\prime}}{\omega-g},
\end{equation}
The derivative with respect to $g^{\prime}$ and the mixed derivatives are
given by the expressions%
\begin{align}
\partial_{g^{\prime}}^{2}\tilde{R}  &  =-\frac{\omega^{2}+g^{\prime2}%
}{4\left(  \omega^{2}-g^{\prime2}\right)  ^{2}}+\frac{1}{8}\left[
\frac{\kappa_{+}^{-1}}{\nu_{+}^{2}-1}-\frac{\kappa_{-}^{-1}}{\nu_{-}^{2}%
-1}\right] \nonumber\\
&  +\frac{1}{8}\left[  \frac{f(\nu_{+}^{-1})}{\sqrt{\kappa_{+}}(\omega
+g^{\prime})}-\frac{f(\nu_{-}^{-1})}{\sqrt{\kappa_{-}}(\omega-g^{\prime}%
)}\right] \\
\partial_{g,g^{\prime}}^{2}\tilde{R}  &  =\partial_{g^{\prime},g}^{2}\tilde
{R}\nonumber\\
&  =\frac{1}{8}\left[  \frac{1}{\nu_{+}^{2}-1}-\frac{1}{\nu_{-}^{2}-1}%
+\frac{f(\nu_{+}^{-1})}{\nu_{+}}-\frac{f(\nu_{-}^{-1})}{\nu_{-}}\right]  ,
\end{align}
which evaluated in $P_{0}$, give
\begin{align}
\partial_{g}^{2}\tilde{R}  &  =\partial_{g^{\prime}}^{2}\tilde{R}=\frac{1}%
{4}\left(  \frac{1}{\omega^{2}(\omega^{2}-1)}+\omega^{-1}f(\omega
^{-1})\right)  ,\label{d2RP0}\\
\partial_{g,g^{\prime}}^{2}\tilde{R}  &  =\partial_{g^{\prime},g}^{2}\tilde
{R}=\frac{1}{4}\left(  \frac{1}{\omega^{2}-1}-\omega^{-1}f(\omega
^{-1})\right)  . \label{d2RpP0}%
\end{align}
We then compute the determinant of the Hessian in $P_{0}$, obtaining the
following expression%
\begin{align}
\det H  &  =\partial_{g}^{2}\tilde{R}\times\partial_{g^{\prime}}^{2}\tilde
{R}-(\partial_{g,g^{\prime}}^{2}\tilde{R})^{2}\nonumber\\
&  =\frac{\left(  \omega^{2}+1\right)  \left(  2\omega f(\omega^{-1}%
)-1\right)  }{16\omega^{4}\left(  \omega^{2}-1\right)  }, \label{HESSIAN}%
\end{align}
which is always positive because%
\begin{equation}
f(\omega^{-1})>\frac{1}{\omega},\text{ for }\omega\geq1~.
\label{CONDITION-on-F1}%
\end{equation}
We have also checked that $\det H>0$ in the limit of $\omega\rightarrow1^{+}$.
Finally we have verified that the second-order derivative of Eq.~(\ref{d2RP0})
is positive in $P_{0}$. In fact, for $\omega>1$, one always has%
\begin{equation}
\frac{1}{4}\left(  \frac{1}{\omega^{2}(\omega^{2}-1)}+\omega^{-1}f(\omega
^{-1})\right)  >0.
\end{equation}
Therefore, $P_{0}$ is a point of absolute minimum for the key-rate of
Eq.~(\ref{FORMULA-RATE-HET-HOM}), so that Eq.~(\ref{R-Rcoll}) is also verified
for the switching protocol.

\end{document}